\documentclass[12pt]{article}
\usepackage{moreverb}
\usepackage[left=1in,right=1in,top=1in,bottom=1in]{geometry}
\usepackage{graphicx}
\usepackage{booktabs}
\usepackage{setspace}
\usepackage{multirow}
\usepackage[]{lineno}
\usepackage{amsmath}
\usepackage{amsmath}
\usepackage{amssymb}
\usepackage{enumerate}
\usepackage{xcolor}
\usepackage{color,colortbl}
\usepackage[colorlinks=true, linkcolor= red,linkbordercolor=green,citecolor=blue,urlcolor=blue]{hyperref}
\usepackage[authoryear,round]{natbib}%
\usepackage{lineno}
\usepackage{float,lscape}
\usepackage{array,longtable,soul}
\usepackage[utf8]{inputenc}
\usepackage{amsmath}
\usepackage{algorithm}
\usepackage{algpseudocode}
\usepackage{subcaption}
\usepackage{float}
\usepackage{placeins}
\usepackage[english]{babel}
\usepackage{blindtext,changepage}
\usepackage{amsfonts}
\usepackage{tikz}
\usetikzlibrary{shapes.geometric, arrows, positioning, calc}
\usepackage{color} 
\definecolor{mygreen}{RGB}{28,172,0} 
\definecolor{mylilas}{RGB}{170,55,241}

\usepackage[most,listings]{tcolorbox}
\tcbset{
	human/.style={colback=orange!5!white, colframe=orange, fonttitle=\bfseries},
	assistant/.style={colback=gray!5!white, colframe=gray, fonttitle=\bfseries},
	code/.style={colback=gray!10!white, colframe=gray!50!black, fonttitle=\bfseries, breakable}
}

\author{\normalsize{Atma Sharma\textsuperscript{1a}, Jie Zhang\textsuperscript{1b}, Meng Lu\textsuperscript{1c},} \normalsize Shuangyi Wu\textsuperscript{1d}, and \normalsize Baoxiang Li\textsuperscript{2} \vspace{+0.50ex}%
 }

\date{ 	 \small{$^{1}$ Key Laboratory of Geotechnical and Underground Engineering of Ministry of Education and Department of Geotechnical Engineering, Tongji University, Shanghai, China.} \\ \vspace{2pt}  \small{$^{2}$ SenseTime Research, Beijing, China.  \\ \vspace{0.25cm}  $^{a}$ \url{atmasharma@tongji.edu.cn}, $^{b}$ \url{cezhangjie@tongji.edu.cn},  $^{c}$ \url{lumeng@tongji.edu.cn}, \\$^{d}$ \url{sywu@tongji.edu.cn}, $^{2}$  \url{libaoxiang@sensetime.com}}}   

\title{ \large {\textnormal{Programming Geotechnical Reliability Algorithms using Generative AI}}}

\usepackage{titling}
\newgeometry{margin=1in}

\let\oldabstract\abstract
\let\oldendabstract\endabstract
\makeatletter
\renewenvironment{abstract}
{%
	{\list{}{\addtolength{\leftmargin}{-2.5em} 
			\listparindent 1.5em%
			\itemindent    \listparindent%
			\rightmargin   \leftmargin%
			\parsep        \z@ \@plus\p@}%
		\item\relax}%
	{\endlist}%
	\oldabstract}
{\oldendabstract}
\makeatother

\begin{document}
	
	\begin{onehalfspace}
		\maketitle	
		
	\end{onehalfspace}
	
	\vspace{-0.75cm}
	\doublespacing
	\begin{abstract}
		\noindent\rule{16.5cm}{1pt}
Programming reliability algorithms is crucial for risk assessment in geotechnical engineering.  This study explores the possibility of automating and accelerating this task using Generative AI based on Large Language Models (LLMs). Specifically, the most popular LLM, i.e., ChatGPT, is used to test the ability to generate MATLAB codes for four classical reliability algorithms. The four specific examples considered in this study are: (1) First Order Reliability Method (FORM);  (2) Subset simulation; (3) Random field simulation; and  (4) Bayesian update using Gibbs sampling. The results obtained using the generated codes are compared with benchmark methods. It is found that the use of LLMs can be promising for generating reliability codes. Failure, limitations, and challenges of adopting LLMs are also discussed. Overall, this study demonstrates that existing LLMs can be leveraged powerfully and can contribute toward accelerating the adoption of reliability techniques in routine geotechnical engineering. 

		\noindent\rule{16.5cm}{1pt} 
	\end{abstract}
	\clearpage
	\section*{Introduction}\label{Section : Introduction}
	
	Reliability assessment is one of the fundamental task in civil and environmental engineering \citep[e.g.,][]{christian2004geotechnical,ang2007probability,fenton2008risk,phoon2015risk,ayyub2016probability,zhang2023geotechnical}. Due to inevitable uncertainties in geological (e.g., soil/rock properties \citep[e.g.,][]{sharma2022hierarchical,sharma2023spectral}) and environmental (e.g., rainfall intensity \citep[e.g.,][]{sharma2024constructing,lu2024stochastic}) conditions, it become important to use statistical and probability techniques for quantifying the reliability. Since a majority of conventional engineers and practioners are not usually trained in statistics and probability, using such techniques can become challenging in end use practitioners. For such reasons, the adoption of reliability technique is routine or large scale projects is still not quite popular \citep{phoon2022geotechnical} across some civil engineering disciplines (e.g., geotechnical engineering). As infrastructure projects grow in scale and complexity, in the current and upcoming wake of climate change and its associated natural hazards, and in the context of urban resilience, the need for sophisticated risk assessment has become increasingly critical \citep[e.g.,][]{leichenko2011climate}. 
	
	In recent years, the advent of large language models (LLMs) has opened up new possibilities for enhancing the development of codes \citep{Brown2020}. Large language models, such as GPT-3 and its successors, are artificial intelligence systems trained on vast amounts of text data, enabling them to understand and generate human-like text across a wide range of domains \citep{OpenAI2020}. While initially developed for natural language processing tasks, these models have demonstrated remarkable capabilities in code generation and comprehension \citep{Chen2021}. One of the most promising aspects of using large language models in reliability coding is their potential to bridge the gap between theoretical concepts and practical implementation. Programming reliability algorithms often involves complex mathematical models and statistical methods that can be challenging to translate into efficient code. Large language models, with their ability to understand both natural language descriptions of reliability principles and programming constructs, can serve as intermediaries in this translation process. Engineers can describe their intended reliability analysis approach in plain language, and the LLM can generate corresponding code that implements the described methodology.

Another potential advantage of LLMs is the potential for accelerated reliability software development. Traditional software development in this domain often requires specialized knowledge of civil engineering principles, reliability principles, as well as programming paradigms, creating a high barrier to entry and slowing the pace of innovation. LLMs can serve as intelligent coding assistants, capable of understanding high-level descriptions of engineering problems and generating corresponding code snippets or even entire algorithms \citep{Austin2021}. This capability allows engineers to focus more on the conceptual aspects of risk assessment and reliability analysis, rather than getting bogged down in the intricacies of syntax and implementation details. Moreover, LLMs can facilitate rapid prototyping and iterative development of risk solution softwares. By quickly generating multiple code variations based on different approaches or parameters, engineers can explore a wider range of solutions and identify optimal strategies more efficiently \citep{Li2022}. This accelerated experimentation process has the potential to drive innovation and improve the overall quality of reliability assessments softwares or environments.

	The capability of programming numerical algorithms was explored in \cite{kashefi2023chatgpt}. Various mathematical problems such as the Poisson equation, the diffusion equation, the incompressible Navier-Stokes equations, compressible inviscid flow, eigenvalue problems, solving linear systems of equations, etc., were tested. \cite{sobania2023analysis} tested the bug fixing performance in codes using ChatGPT. \cite{liu2024your} proposed an evaluation framework for codes generated using ChatGPT .
\cite{prieto2023investigating} utilized chatgpt for scheduling construction projects.
\cite{kim2024chatgpt}   tested the chatgpt ability to generate codes for solving seepage flow and slope stability, and the image processing of X-ray computed tomographic image for partially saturated sand. \cite{kumar2024geotechnical} explored the context-specific search engine and integrating ChatGPT  with engineering workflows and tools to improve reasoning abilities and accuracy for complex tasks.  \cite{chen2024investigation} investigated the capability to answer basic geotechnical engineering textbooks questions using ChatGPT . Recently \cite{wu2024future} explored the possibility of solving advanced geotechnical problem using ChatGPT . Four problems were considered, i.e,  slope stability assessment, microzoning by seismic risk, simulation parameter recommendation and site similarity prediction

Given the promising use of ChatGPT in aforementioned studies, the objective of this study is to investigate the possibility of generating codes specific to risk and reliability. The authors stick to ChatGPT (not some other models e.g., Claude, Gemini, Grok, and Llama) as the choice of generative AI mainly because it is already quite popular. Four classical reliability methods are considered for investigation: (1) First Order Reliability Method (FORM);  (2) Subset simulation; (3) Random field simulation; and  (4) Bayesian updating. The results obtained using the generated codes are compared with benchmark methods. The results seem promising and can be used successfully in future for testing further algorithms related to reliability.  Failure, limitations, and challenges of adopting LLMs are also discussed. Overall, this study demonstrates that existing LLM based Generative AI can be leveraged powerfully and can contribute toward accelerating the adoption of reliability techniques in routine civil engineering. 

The rest of this manuscript is organized as discussed next. The four reliability examples are considered one by one. The generated code in response to query is presented for each example and the obtained results are bechmarked against an independent method. Finally the challenges, limitations, and future scope are discussed.

\section*{Prompt Engineering for Generative AI Models}

Recent advancements in generative AI models, such as OpenAI’s GPT-4 \citep{OpenAI2020}, have introduced powerful tools for automating complex tasks. A critical aspect of using these models effectively is prompt engineering, which involves carefully crafting inputs to obtain accurate and relevant outputs. In this paper, we primarily employ zero-shot prompting, but it is important to understand the broader spectrum of prompting techniques, as they offer varying degrees of control over AI responses. Prompt engineering can be categorized into several key approaches, including zero-shot prompting, few-shot prompting, and chain-of-thought prompting. 

Zero-shot prompting \citep[e.g.,][]{kojima2022large}, relies on the AI model's ability to generate solutions without any prior examples or specific training for the task at hand. This approach is advantageous when dealing with generalizable problems, as it allows the model to draw from its pre-existing knowledge to provide responses. Zero-shot prompting enables the model to implement algorithms based solely on natural language descriptions of the problem, without requiring task-specific examples.

Few-shot prompting \citep[e.g.,][]{reynolds2021prompt} introduces a limited number of examples within the input prompt to guide the AI model toward producing more precise outputs. This method is particularly useful when addressing problems where domain-specific knowledge is required, or when there is a need to demonstrate specific output formats. By providing a few representative examples, the model is better equipped to generalize to similar problems, making it effective for tasks like iterative design optimizations or generating code snippets that conform to specific engineering standards.

Another approach, chain-of-thought prompting \citep[e.g.,][]{wei2022chain}, structures the input to guide the model through intermediate reasoning steps before arriving at a final answer. This method enhances the model's ability to handle complex, multi-step problems by explicitly laying out the reasoning process in the prompt. Chain-of-thought prompting could be beneficial in problems involving layered computations, such as stepwise risk assessments or staged reliability calculations, where the solution requires a sequence of dependent operations.

The reason for choosing zero-shot prompting in this study is because of its simplicity. Zero-shot prompting is the most straightforward approach to interact with generative AI, as it requires no specific examples or complex input structures. This makes it accessible even to non-experts or normal practitioners who may lack technical knowledge of AI. Engineers can focus on describing the problem rather than understanding complex prompt approaches, reducing cognitive load and allowing them to concentrate on domain-specific issues. With no need for additional training, data preparation, or fine-tuning, zero-shot prompting is a cost-effective approach, especially for smaller projects or studies with limited resources.

\section*{Example 1: First Order Reliability Method}

The First Order Reliability Method (FORM) is a widely applied technique in structural reliability analysis for estimating the probability of failure of engineering systems \citep{Hasofer1974}. FORM approximates the limit state function using a first-order Taylor series expansion at the design point, which is the point on the limit state surface closest to the origin in the transformed standard normal space \citep{Rackwitz1978}. The basic idea of FORM can be summarized as follows:

\begin{enumerate}
	\item \textnormal{Transformation of Random Variables to Standard Normal Space:} \\
	The initial step in FORM involves transforming the original basic random variables $\mathbf{X} = [X_1, X_2, \dots, X_n]^\top$ into standard normal variables $\mathbf{U} = [U_1, U_2, \dots, U_n]^\top$ using \citep{Ditlevsen1996}:
\begin{equation}
	U_i = \Phi^{-1}\left[F_{X_i}(x_i)\right], \quad i = 1, 2, \dots, n
\end{equation}
	where $\Phi^{-1}(\cdot)$ denotes the inverse of the cumulative distribution function (CDF) of the standard normal distribution, and $F_{X_i}(x_i)$ is the CDF of the random variable $X_i$. 
	
	\item \textnormal{Definition of the Limit State Function in $U$-Space:} \\
	The limit state function is expressed as:
\begin{equation}
	g(\mathbf{U}) = 0
\end{equation}
	where $g(\mathbf{U})$ separates the safe domain $g(\mathbf{U}) > 0$ from the failure domain $g(\mathbf{U}) < 0$ \citep{Melchers2018}. 
	
	\item \textnormal{Identification of the Design Point $\mathbf{u}^*$:} \\
	The design point, also known as the most probable point (MPP), is the point on the limit state surface $g(\mathbf{U}) = 0$ closest to the origin. It is obtained by solving the following constrained optimization problem \citep{Liu2019}:
\begin{equation}
	\mathbf{u}^* = \arg\min_{\mathbf{u}} \left\{ \|\mathbf{u}\| : g(\mathbf{u}) = 0 \right\}
\end{equation}
	where $\|\mathbf{u}\| = \sqrt{u_1^2 + u_2^2 + \dots + u_n^2}$ represents the Euclidean norm of $\mathbf{u}$.
	
	\item \textnormal{Calculation of the Reliability Index $\beta$:} \\
	The reliability index $\beta$ is defined as the distance from the origin to the design point in the $U$-space \citep{Hasofer1974}:
\begin{equation}
	\beta = \|\mathbf{u}^*\|
\end{equation}
	\item \textnormal{Estimation of the Probability of Failure $P_f$:} \\
	The probability of failure is approximated as \citep{Rackwitz1978}:
\begin{equation}
	P_f \approx \Phi(-\beta)
\end{equation}
	where $\Phi(\cdot)$ is the CDF of the standard normal distribution.
\end{enumerate}

FORM is computationally efficient and well-suited for problems involving a small number of random variables and limit state functions that are either linear or moderately nonlinear \citep{Melchers2018}. However, FORM may provide less accurate results for highly nonlinear limit state functions, systems with multiple design points, or when the failure surface has significant curvature near the design point \citep{DerKiureghian2005}. 

\begin{figure}[t]
	\centering
	\includegraphics[trim=0cm 0cm 0.0cm 0.0cm, clip=true,width=1\textwidth]{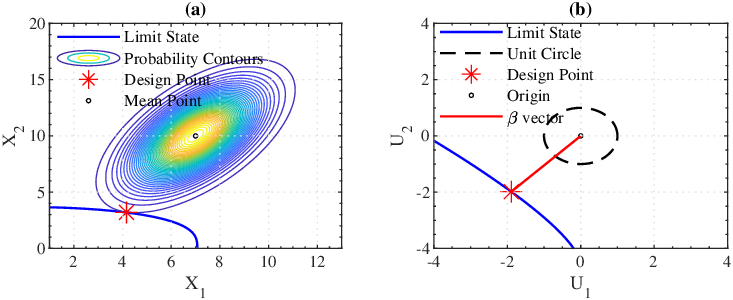}
	\caption{Results obtained using the code generated by Chat-GPT for FORM in: (a) $X$ space; and (b) $U$ space.}
	\label{FORM_1}
\end{figure}

\begin{figure}[t]
	\centering
	\includegraphics[trim=0cm 0cm 0.0cm 0.0cm, clip=true,width=1\textwidth]{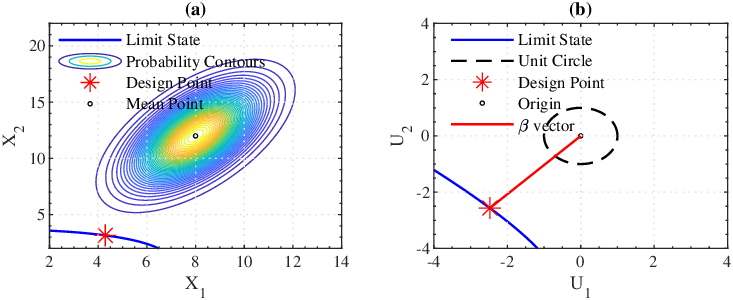}
	\caption{Results obtained using the code generated by Chat-GPT for FORM in: (a) $X$ space; and (b) $U$ space.}
	\label{FORM_2}
\end{figure}

Now, let us ask ChatGPT to write a MATLAB code for the First Order Reliability Method (FORM). The query and response are presented in Listing \ref{lst:label}. The implementation appears to be correct. The results obtained using the code in Listing \ref{lst:label} are shown in Fig. \ref{FORM_1}. The calculated reliability index is $\beta = 2.74$, and the probability of failure is $P_f = 3.07 \times 10^{-3}$, achieved after 6 iterations. To confirm these results, standard Monte Carlo Simulation (MCS) with a sample size of $10^7$ was performed, yielding $P_f = 2.87 \times 10^{-3}$ and $\beta = 2.76$. The results are in close agreement. To further verify the findings, the mean values of the random variables were set to $\mu = [8,12]$, while the rest of the code remained unchanged. The revised results are shown in Fig. \ref{FORM_2}. The reliability index obtained is $\beta = 3.56$ and the probability of failure is $P_f = 1.83 \times 10^{-4}$, requiring 7 iterations. The MCS results with a sample size of $10^7$ are $P_f = 1.73 \times 10^{-4}$ and $\beta = 3.58$. Based on the results presented in Fig. \ref{FORM_1} and Fig. \ref{FORM_2}, the results obtained using FORM seem accurate and consistent.

\begin{samepage}

	\begin{tcolorbox}[human, title=Human:]
Write me  a MATLAB code for solving a reliability problem using Advanced first order reliability method. AFORM is basically a constrained optimization problem. Also use correlated random variables and non-linear limit state function.
	\end{tcolorbox}
	
	\begin{tcolorbox}[assistant, title=AI Assistant:]
Certainly! I'll create a MATLAB script for solving a reliability problem using the Advanced First Order Reliability Method (AFORM) with correlated random variables and a nonlinear limit state function. Here's the code:
	\end{tcolorbox}
	
	\begin{tcolorbox}[code, title=Matlab Script]
		\begin{lstlisting}[caption={First Order Relaibility Method},label={lst:label},lineskip=-2pt,language=Matlab]
clc; clear all; close all;
% AFORM for Reliability Problem with Correlated Variables
% Define problem parameters
n = 2; % Number of random variables
mu = [7; 10]; % Mean values of random variables
sigma = [1.5; 2.5]; % Standard deviations of random variables

% Correlation matrix
rho = [1.0 0.6;
0.6 1.0];

% Cholesky decomposition of correlation matrix
L = chol(rho, 'lower');

% Define a non-linear limit state function with polynomial terms
g = @(x) x(1)^2 + x(2)^3 - 50; % Example non-linear limit state function

% AFORM algorithm parameters
max_iter = 100;
tolerance = 1e-6;

% Initialize
x = mu;
u = zeros(n, 1);
beta = inf;
iter = 0;

while iter < max_iter
% Transform from X-space to correlated U-space
u_corr = (x - mu) ./ sigma;

% Transform to uncorrelated U-space
u = L \ u_corr;

% Evaluate limit state function and its gradient
G = g(x);
dG = zeros(n, 1);
h = 1e-5; % Step size for numerical gradient
for i = 1:n
x_plus = x;
x_plus(i) = x(i) + h;
dG(i) = (g(x_plus) - G) / h;
end

% Transform gradient to uncorrelated U-space
dG_u = L' * (dG .* sigma);

% Compute alpha vector
alpha = -dG_u / norm(dG_u);

% Compute new beta
beta_new = -G / (dG_u' * alpha) + u' * alpha;

% Check convergence
if abs(beta_new - beta) < tolerance || norm(beta_new * alpha - u) < tolerance
break;
end

% Update beta and design point
beta = beta_new;
u = beta * alpha;
u_corr = L * u;
x = mu + sigma .* u_corr;
iter = iter + 1;
end

% Compute probability of failure
Pf = normcdf(-beta);

% Display results
fprintf('Correlated AFORM Results:\n');
fprintf('Reliability index (beta): %.4f\n', beta);
fprintf('Probability of failure: %.4e\n', Pf);
fprintf('Design point in X-space:\n');
disp(x);
fprintf('Design point in uncorrelated U-space:\n');
disp(u);
fprintf('Number of iterations: %d\n', iter);


			
		\end{lstlisting}
	\end{tcolorbox}
\end{samepage}

\section*{Example 2: Subset simulation}

Subset simulation is an advanced Monte Carlo method designed for estimating small failure probabilities in complex systems \citep{au2001estimation}. Unlike traditional MCS, which requires an large number of samples to estimate rare events, Subset simulation decomposes the failure probability into a product of larger, more easily estimated conditional probabilities. The failure probability $P_f$ is expressed as:

\begin{equation}
	P_f = P(F_1) \prod_{i=2}^{m} P(F_i \mid F_{i-1}),
\end{equation}

where $F_1 \supset F_2 \supset \cdots \supset F_m = F$ are a sequence of nested intermediate failure events leading to the final failure event $F$. The basic idea of the subset simulation can be summarized as follows.

\begin{enumerate}
	\item Generate $N$ samples from the original distribution and estimate the first conditional probability $P(F_1)$.
	\item For each subsequent intermediate failure event $i = 2, \ldots, m$:
	\begin{itemize}
		\item Use Markov Chain Monte Carlo (MCMC) sampling to generate samples conditional on $F_{i-1}$.
		\item Estimate $P(F_i \mid F_{i-1})$ using these samples.
	\end{itemize}
	\item Compute the overall failure probability $P_f$ as the product of the estimated probabilities.
\end{enumerate}

A common approach for defining the intermediate failure events is to set:

\begin{equation}
	F_i = \{ g(\mathbf{X}) \leq y_i \},
\end{equation}

where $y_1 > y_2 > \cdots > y_m = 0$, and $y_i$ are threshold levels selected such that $P(F_i \mid F_{i-1}) \approx p_0$ for a chosen constant $p_0$ (typically $p_0 = 0.1$). This ensures that each conditional probability remains large enough for efficient estimation. The Modified Metropolis-Hastings algorithm is frequently employed for the MCMC sampling step. For each component $j$ of the random vector $\mathbf{X}$, a candidate sample is generated as follows:

\begin{equation}
	\xi_j^{(k)} = \xi_j^{(k-1)} + \sigma_j Z_j,
\end{equation}

where $Z_j \sim N(0, 1)$ is a standard normal random variable, and $\sigma_j$ is the standard deviation of the proposal distribution. The candidate $\boldsymbol{\xi}^{(k)}$ is accepted with probability:

\begin{equation}
	\alpha = \min \left\{ 1, \frac{f_X(\boldsymbol{\xi}^{(k)})}{f_X(\boldsymbol{\xi}^{(k-1)})} \right\},
\end{equation}

where $f_X(\boldsymbol{\xi})$ represents the probability density function of the original distribution. Subset Simulation is particularly effective for problems involving small failure probabilities and high-dimensional random variables, making it widely used in structural reliability analysis and geotechnical engineering applications.

\begin{figure}[t]
	\centering
	\includegraphics[trim=0cm 0cm 0.0cm 0.0cm, clip=true,width=1\textwidth]{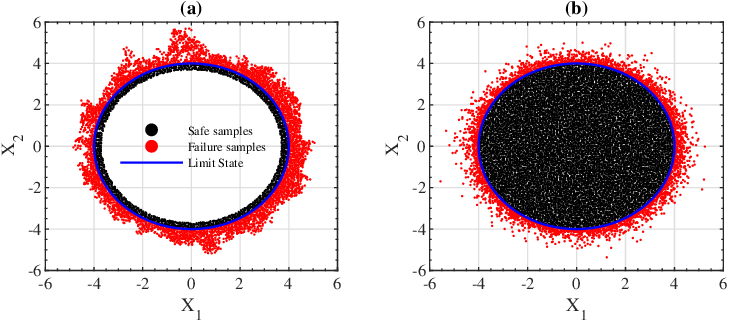}
	\caption{Results obtained using the code generated by ChatGPT for: (a) Subset Simulation on a  circular limit state function ($P_f=3.14\times 10^{-4}$); and (b) using MCS with sample size $10^8$ ($P_f=3.36\times 10^{-4}$).}
	\label{SS_2}
\end{figure}

\begin{figure}[t]
	\centering
	\includegraphics[trim=0cm 0cm 0.0cm 0.0cm, clip=true,width=0.65\textwidth]{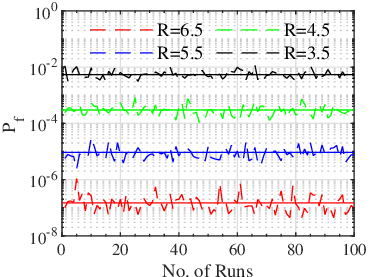}
	\caption{Results obtained using the code generated by ChatGPT for Subset simulation on a  circular limit state function. The horizontal solid line at each R denotes the results obtained using crude Monte Carlo Samples with sample size $10^8$.}
	\label{SS_1}
\end{figure}

Now, let us ask ChatGPT to write a MATLAB code for the Subset Simulation. Listing \ref{2nd:label} presents the query and response. For the current example, we have set correlation coeffecient to $0$ (originally, it was 0.5 in the response by ChatGPT). The next example will show a correlated case for a different limit state. The results obtained using the code in Listing \ref{2nd:label} are shown in Fig. \ref{SS_2}. The calculated $P_f = 3.14 \times 10^{-3}$. To confirm these results, standard MCS with a sample size of $10^8$ was performed, yielding $P_f = 3.36 \times 10^{-3}$. The results are close enough. To further verify the findings, the radius $R$ of the circular limit state was changed. 4 scenarios are considered with $R=[3.5,4.5,5.5,6.5]$.  A total of 100 runs were performed using the code provided in Listing \ref{2nd:label}. The results are presented in Fig. \ref{SS_1}. The $P_f$ values make sense as higher $R$ should lead to lower $P_f$ and vice versa. The results are also benchmarked against MCS with $10^8$ samples. Be noted that the subset simulation results and MCS results are in close agreement, indicating that the code generated by ChatGPT is working fine. 

\begin{figure}[t]
	\centering
	\includegraphics[trim=0cm 0cm 0.0cm 0.0cm, clip=true,width=1\textwidth]{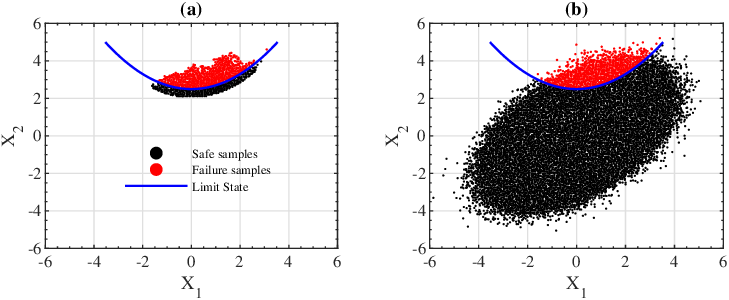}
	\caption{Results obtained using the code generated by ChatGPT for: (a) Subset Simulation on a  parabolic limit state function ($P_f=2.35\times 10^{-3}$); and (b) using MCS with sample size $10^8$ ($P_f=2.36\times 10^{-3}$).}
	\label{SS_3}
\end{figure}

The previous two examples were on a circular limit state and with uncorrelated random variables. We did not provide any input about the choice of limit state in the prompt in Listing \ref{2nd:label}. ChatGPT chooses a circular limit state. Now, let's change the circular limit state to parabolic and change the correlation coefficient to 0.5. The obtained subset simulation results, along with crude MCS results, are presented in Fig. \ref{SS_3}. The results seem to be okay in comparison to MCS. Based on the results in Figs. \ref{SS_2}, Fig. \ref{SS_1}, and Fig. \ref{SS_3}, it seems that the code is working fine.

\begin{samepage}

	\begin{tcolorbox}[human, title=Human:]
do you know about subset simulation, it a technique to compute failure probability when it is too small, usually crude MCS can require too many samples in case of low failure probability, if you know it, please provide me an example with MATLAB code
	\end{tcolorbox}
	
	\begin{tcolorbox}[assistant, title=AI Assistant:]
Yes, I'm familiar with subset simulation. It's an advanced Monte Carlo method particularly useful for estimating small failure probabilities in reliability analysis, especially when crude Monte Carlo simulation would require an impractically large number of samples.
I'd be happy to provide you with a MATLAB example of subset simulation. This technique uses a sequence of intermediate failure events to estimate the probability of the target failure event, which is particularly efficient for rare events.
	\end{tcolorbox}
	
	\begin{tcolorbox}[code, title=Matlab Script]
		\begin{lstlisting}[caption={Subset Simulation},label={2nd:label},lineskip=-2pt,language=Matlab]
% Subset Simulation for Reliability Analysis
clc; clear all; close all;
% Problem setup
n = 2;  % Number of random variables
mu = [0; 0];  % Mean values
sigma = [1; 1];  % Standard deviations
rho =0;  % Correlation coefficient

% Correlation matrix
Corr = [1 rho; rho 1];
% Cholesky decomposition of correlation matrix
L = chol(Corr, 'lower');

% Circular limit state function
R = 4.0;  % Radius of failure surface
g = @(x) R - sqrt(sum(x.^2));  % Failure when g(x) <= 0

% Subset simulation parameters
N = 20000;  % Number of samples per subset
p0 = 0.1;  % Probability of each subset
Ns = ceil(p0 * N);  % Number of Markov Chain samples

% Initialize
max_levels = 20;
Z = randn(N, n);  % Standard normal random variables
X = repmat(mu', N, 1) + (Z * L') .* repmat(sigma', N, 1);  % Correlated random variables
F = arrayfun(@(i) g(X(i,:)'), 1:N)';

% Main subset simulation loop
for level = 1:max_levels
% Sort samples based on limit state function values
[F_sorted, idx] = sort(F, 'ascend');

% Threshold for this level
threshold = F_sorted(Ns);

% Check if target failure probability is reached
if threshold <= 0
break;
end

% Select seeds for MCMC
X_seeds = X(idx(1:Ns), :);

% Generate new samples using MCMC
X_new = zeros(N, n);
F_new = zeros(N, 1);

for i = 1:N
if i <= Ns
X_new(i,:) = X_seeds(i,:);
F_new(i) = F_sorted(i);
else
% Modified Metropolis-Hastings algorithm
current = X_new(i-1, :);
proposal_std = 0.1;  % Proposal standard deviation

% Generate correlated proposal
Z_proposal = randn(1, n);
proposal = current + (Z_proposal * L') * proposal_std;

F_current = g(current');
F_proposal = g(proposal');

if F_proposal <= threshold
% Calculate multivariate normal PDFs
p_current = mvnpdf(current, mu', L*diag(sigma)*L');
p_proposal = mvnpdf(proposal, mu', L*diag(sigma)*L');
alpha = min(1, p_proposal / p_current);
else
alpha = 0;
end

if rand() < alpha
X_new(i,:) = proposal;
F_new(i) = F_proposal;
else
X_new(i,:) = current;
F_new(i) = F_current;
end
end
end

X = X_new;
F = F_new;
end

% Compute failure probability using Subset Simulation
Pf_SS = p0^(level-1) * sum(F <= 0) / N;
disp(['Estimated Probability of Failure (Subset Simulation): Pf_SS = ', num2str(Pf_SS)]);

			
		\end{lstlisting}
	\end{tcolorbox}
\end{samepage}

\section*{Example 3: Random field simulation}

Random field simulation becomes important in civil engineering \citep{fenton2008risk} for capturing the inherent spatial variability of soil properties or material properties. A random field \( Z(\mathbf{x}) \) can be described by its mean function \( \mu(\mathbf{x}) \), standard deviation function \( \sigma(\mathbf{x}) \), and spatial correlation structure \( \rho(\tau) \). The general form of a random field is given by:


\begin{equation}
	Z(\mathbf{x}) = \mu(\mathbf{x}) + \sigma(\mathbf{x}) Y(\mathbf{x})
\end{equation}

where \( Y(\mathbf{x}) \) is a standardized random field with zero mean and unit variance. The standardized field \( Y(\mathbf{x}) \) is typically modeled to have a specific spatial correlation structure. The spatial correlation structure of the random field is defined by the correlation function \( \rho(\tau) \), where \( \tau = \lVert \mathbf{x}_1 - \mathbf{x}_2 \rVert \) is the lag distance between two points \( \mathbf{x}_1 \) and \( \mathbf{x}_2 \) in the domain. A commonly used correlation function is the \textit{exponential correlation function}:


\begin{equation}
	\rho(\tau) = \exp\left(-\frac{2\tau}{\theta}\right)
\end{equation}

where \( \theta \) represents the correlation length, which controls the extent of spatial dependency.

Several methods exist to generate realizations of random fields, each with varying computational complexity and accuracy. Four common approaches are:

\subsubsection*{Karhunen-Loève Expansion (KLE)}

The Karhunen-Loève Expansion \citep[e.g.,][]{schwab2006karhunen} is a spectral method that expresses the random field as a series of orthogonal eigenfunctions:


\begin{equation}
	Z(\mathbf{x}) \approx \mu(\mathbf{x}) + \sigma(\mathbf{x}) \sum_{i=1}^n \sqrt{\lambda_i} \xi_i \phi_i(\mathbf{x})
\end{equation}

where \( \lambda_i \) and \( \phi_i(\mathbf{x}) \) are the eigenvalues and eigenfunctions of the covariance function \( C(\mathbf{x}_1, \mathbf{x}_2) \) and \( \xi_i \sim \mathcal{N}(0,1) \) are independent standard normal variables. The eigenfunctions \( \phi_i(\mathbf{x}) \) capture the dominant spatial patterns, while the eigenvalues \( \lambda_i \) indicate their relative significance.

\subsubsection*{Covariance Matrix Decomposition}

The covariance matrix decomposition method \citep[e.g.,][]{vanmarcke1986random} generates a random field by decomposing the covariance matrix \( \mathbf{C} \) into the product of a lower triangular matrix \( \mathbf{L} \) and its transpose, using Cholesky decomposition:


\begin{equation}
	\mathbf{C} = \mathbf{L} \mathbf{L}^\top
\end{equation}

The random field can then be generated as:


\begin{equation}
	\mathbf{Z} = \mu + \mathbf{L} \mathbf{Y}
\end{equation}

where \( \mathbf{Y} \sim \mathcal{N}(0, \mathbf{I}) \) is a vector of uncorrelated standard normal variables. This method ensures that the generated field has the desired covariance structure.

\subsubsection*{Circulant Embedding with Fast Fourier Transform (FFT)}

The Circulant Embedding method \citep[e.g.,][]{graham2018analysis} leverages FFT to efficiently generate large random fields by embedding the covariance matrix in a circulant matrix:

\begin{enumerate}
	\item Given the covariance function \( C(\mathbf{x}_1, \mathbf{x}_2) \), a circulant matrix \( \mathbf{C} \) is constructed.
	\item The FFT is applied to \( \mathbf{C} \) to generate uncorrelated Gaussian random variables.
	\item The resulting field is mapped back using the inverse FFT.
\end{enumerate}

This method is computationally efficient, especially for large grids, due to the speed of FFT.

\begin{figure}[t]
	\centering
	\includegraphics[trim=0cm 0cm 0.0cm 0.0cm, clip=true,width=0.75\textwidth]{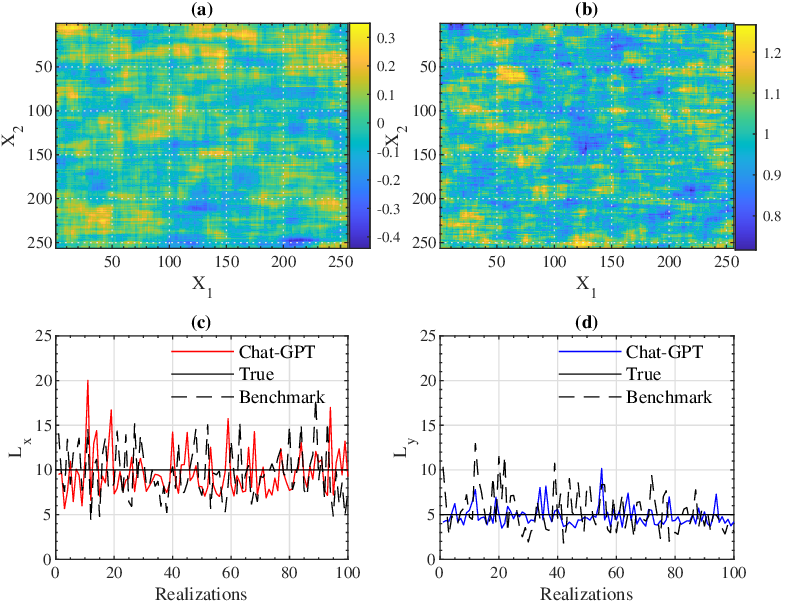}
\caption{Results obtained using the code generated by ChatGPT for 2D random field generation: (a) mean; (b) standard deviation; (c) estimated correlation length $L_x$; and (d) estimated correlation length $L_y$.}
	\label{RF_1}
\end{figure}

\begin{figure}[t]
	\centering
	\includegraphics[trim=0cm 0cm 0.0cm 0.0cm, clip=true,width=0.75\textwidth]{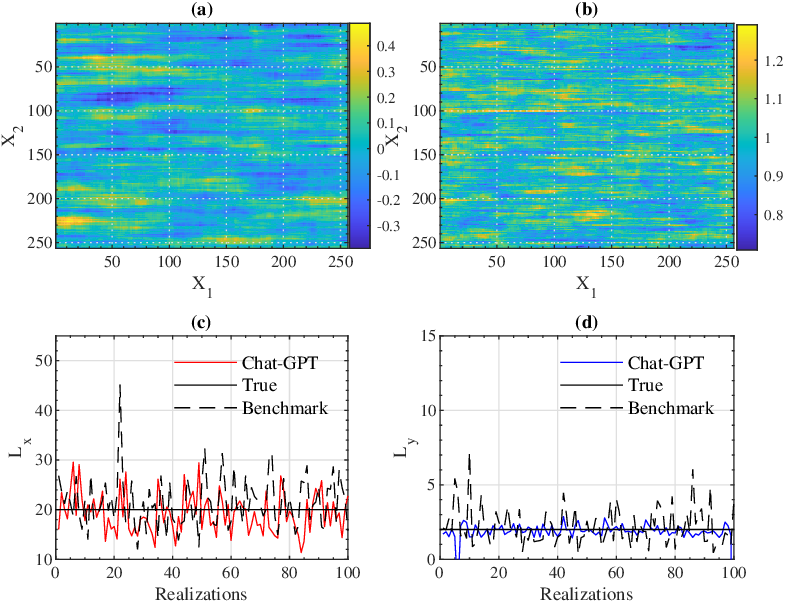}
\caption{Results obtained using the code generated by ChatGPT for 2D random field generation: (a) mean; (b) standard deviation; (c) estimated correlation length $L_x$; and (d) estimated correlation length $L_y$.}
	\label{RF_2}
\end{figure}

Now, let us ask the ChatGPT to write a MATLAB code for random field simulation. A 2D random field example is taken for the current example. Listing \ref{3rd:label} presents the query and response. The code seems to be correct at first glance, but now let us plot the obtained results in Fig. \ref{RF_1}. A total of 100 realizations are generated. Fig. \ref{RF_1}(a-b) presents the mean and standard deviation, respectively. The generated random field seems to be a random field with visual inspection with zero mean and unit variance (see the colobar in Fig. \ref{RF_1}). Also, the figure suggests higher correlation length in x direction. To confirm the results,  the back-obtained correlation length along with the true (prescribed) value are presented in Fig. \ref{RF_1}(c-d). The results are also compared against an independent method using a stepwise covariance matrix (available in \cite{zhang2023geotechnical}; see page no 211 in the book). The results in Fig. \ref{RF_1} indicate that the code can generate realistic random fields with specified correlation lengths. Now, let us check for another example; for this example, we change the correlation length in the x and y direction to 20 and 2, respectively. A similar exercise as in Fig. \ref{RF_1} is conducted, and the results are presented in Fig. \ref{RF_2}. It can be noted that the results are promising.

\begin{samepage}

	\begin{tcolorbox}[human, title=Human:]
Write me an example matlab code for generation of 2D random field with exponential correlation using covariance matrix decompostion method where correlation length in both x and y are seperable 
	\end{tcolorbox}
	
	\begin{tcolorbox}[assistant, title=AI Assistant:]

Sure! Below is the MATLAB code for generating a 2D random field using the Covariance Matrix Decomposition method with an exponential correlation function. The code assumes that the correlation lengths in the x and y directions are separable, meaning the covariance function can be represented as the product of one-dimensional covariance functions in each direction.

	\end{tcolorbox}
	
	\begin{tcolorbox}[code, title=Matlab Script]
		\begin{lstlisting}[caption={Random Field Simulation},label={3rd:label},lineskip=-2pt,language=Matlab]
clc; clear all; close all;

% Parameters
nx = 256; % Number of grid points in x direction
ny = 256; % Number of grid points in y direction
Lx = 100; % Domain size in x direction
Ly = 100; % Domain size in y direction
lx = 10; % Correlation length in x direction
ly =5;  % Correlation length in y direction

% Generate grid
dx = Lx / nx;
dy = Ly / ny;
[x, y] = meshgrid(0:dx:Lx-dx, 0:dy:Ly-dy);

% Generate covariance matrices for x and y directions (separable assumption)
Cx = exp(-abs((0:nx-1)' - (0:nx-1)) * dx / lx); % Covariance in x direction
Cy = exp(-abs((0:ny-1)' - (0:ny-1)) * dy / ly); % Covariance in y direction

% Perform Cholesky decomposition on covariance matrices
Lx_decomp = chol(Cx, 'lower'); % Lower triangular decomposition for x
Ly_decomp = chol(Cy, 'lower'); % Lower triangular decomposition for y

% Generate uncorrelated standard normal random field
Z = randn(ny, nx);

% Apply covariance decomposition to generate correlated field
field = Ly_decomp * Z * Lx_decomp';

% Normalize the field to have zero mean and unit variance
field = (field - mean(field(:))) / std(field(:));

% Plot the generated random field
figure(1);
imagesc(x(1,:), y(:,1), field);
colorbar;
title('Generated 2D Random Field (Covariance Decomposition)');
xlabel('x'); ylabel('y');

			
			
		\end{lstlisting}
	\end{tcolorbox}
\end{samepage}

\section*{Example 4: Bayesian updating}

We denote the site-specific data as \( \mathbf{Y} \) and assign a probabilistic model \( M \) with parameters \( \mathbf{\Theta} \) to this data. In the Bayesian framework, the task of obtaining the posterior distribution \( f(\mathbf{\Theta}|\mathbf{Y}, M) \), can be expressed as follows \citep{ang2007probability}:

\begin{equation}
	f(\mathbf{\Theta}|\mathbf{Y},M) = \frac{f(\mathbf{Y}|\mathbf{\Theta},M)f(\mathbf{\Theta}|M)}{f(\mathbf{Y}|M)} = \frac{f(\mathbf{Y}|\mathbf{\Theta},M)f(\mathbf{\Theta}|M)}{\int f(\mathbf{Y}|\mathbf{\Theta},M)f(\mathbf{\Theta}|M)d\mathbf{\Theta}}
	\label{bayes}
\end{equation}

where \( f(\mathbf{Y}|\mathbf{\Theta},M) \) is the likelihood, \( f(\mathbf{\Theta}|M) \) is the prior distribution, and \( f(\mathbf{Y}|M) \) is the evidence. Given a set of models \( M_1, M_2, \dots, M_k \), the evidence can be used to select the most appropriate model. In many situations, such as reliability-based design (RBD), it is also useful to obtain the posterior predictive distribution, \( f(\mathbf{y}_{new}|\mathbf{Y}) \). 

In many cases, such as in reliability-based design (RBD), it is desirable to determine the posterior predictive distribution, \( f(\mathbf{y}_{new}|\mathbf{Y}) \). This distribution can be obtained using the total probability theorem as follows:

\begin{equation}
	f(\mathbf{y}_{new}|\mathbf{Y}) = \int f(\mathbf{y}_{new}|\mathbf{\Theta}) f(\mathbf{\Theta}|\mathbf{Y}) d\mathbf{\Theta}
	\label{total}
\end{equation}

While both Eq. \ref{bayes} and Eq. \ref{total} might seem straightforward, solving them is challenging due to the high-dimensional integrals involved. Simulation-based methods, such as the Metropolis-Hastings sampler, often become infeasible in higher dimensions of \( \mathbf{\Theta} \). Even one of the most widely used and efficient techniques, the Transitional Markov Chain Monte Carlo (TMCMC) sampler \citep{ching2007transitional}, struggles in higher dimensions. Recently, a Gibbs sampler-based method was proposed by \cite{ching2019constructing} to address the "curse of dimensionality" issue associated with simulation-based approaches. The proposed method leverages conjugate distributions within the multivariate normal probabilistic model, which allows for closed-form solutions of the conditional posterior distributions. This feature enables direct sampling from \( f(\mathbf{\Theta}|\mathbf{Y}, M) \) without the need for rejection sampling. Another strategy was recently proposed by \cite{sharma2024bayesian}, where the high dimensional Bayesian update is translated into a series of lower dimensional update tasks using conditional correlation vine. For demonstration, we stick to \cite{ching2019constructing} approach as it has closed-form expressions for conditional posterior densities. 

\begin{figure}[t]
	\centering
	\includegraphics[trim=0cm 0cm 0.0cm 0.0cm, clip=true,width=1\textwidth]{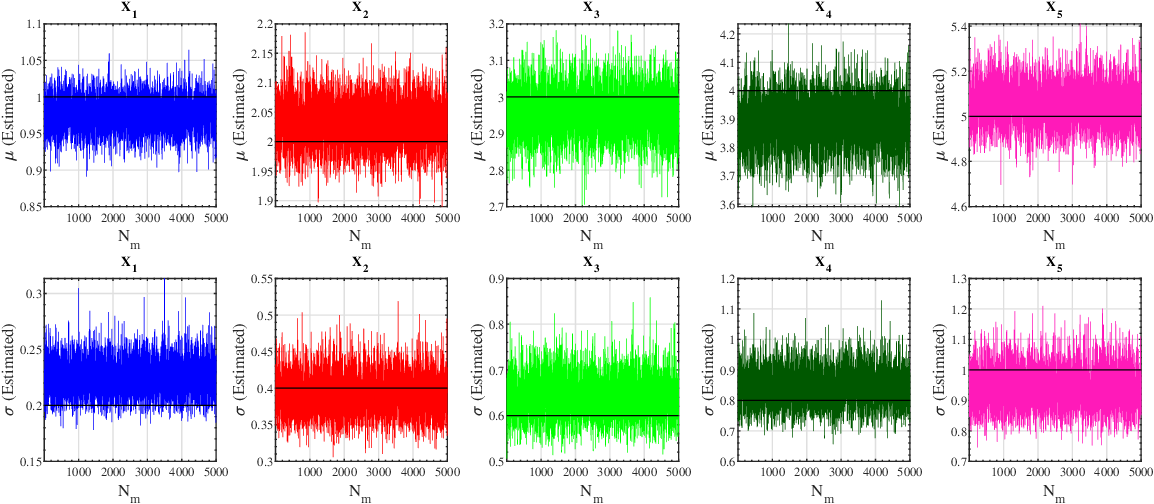}
	\caption{Results obtained using the code generated by Chat-GPT for Bayesian updating on a 5 dimensional data. The colored lines denote the trace plot of the obtained Gibbs samples, and the black horizontal line denotes the true population value.}
	\label{Bayes_2}
\end{figure}

Now, let us ask the ChatGPT to write a MATLAB code for Bayesian updating using Gibbs sampling. Listing \ref{4th:label} presents the query and response. The code appears to be written correctly as it uses iterative Gibbs sampling based on closed-form solutions of conditional densities and iteratively updates the missing values. Now let us check the results obtained using code on 5-dimensional data. The data is generated with $\mu=[1,2,3,4,5]$ in the five dimensions respectively. Coefficient of variation is set to 20 per cent for all the marginals, and random correlation matrices are generated using the vine approach in \cite{sharma2024bayesian} The obtained trace plot of posterior parameters for mean and standard deviation is presented in Fig. \ref{Bayes_2}. The obtained results are in close agreement with the actual value used to generate the samples. This suggested that the code seems to be correct.

\begin{figure}[t]
	\centering
	\includegraphics[trim=0cm 0cm 0.0cm 0.0cm, clip=true,width=0.75\textwidth]{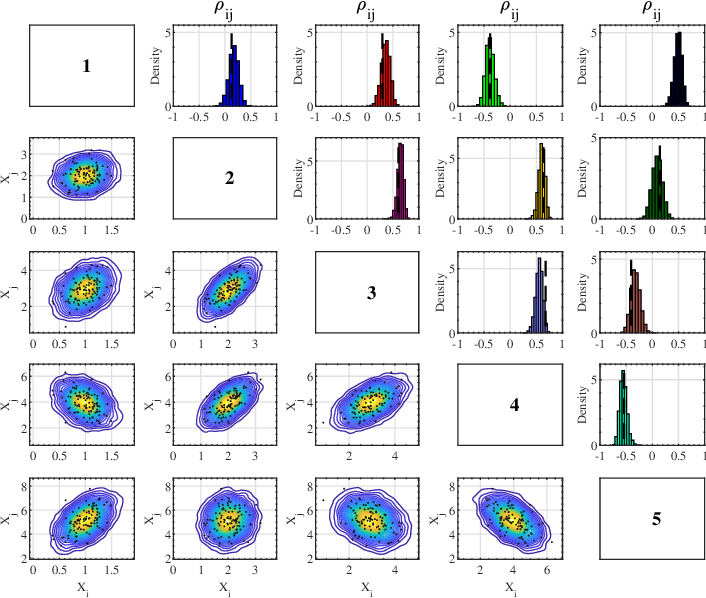}
	\caption{Results obtained using the code generated by Chat-GPT for Bayesian updating on a 5 dimensional data. In the upper triangular matrix, the histogram are for Gibbs samples, and the vertical line denotes the true population value. In the lower triangular matrix, the dots are measured data, and the pairwise empirical contours are obtained from data sampled using the obtained Gibbs samples.}
	\label{Bayes_1}
\end{figure}

\begin{figure}[t]
	\centering
	\includegraphics[trim=0cm 0cm 0.0cm 0.0cm, clip=true,width=1\textwidth]{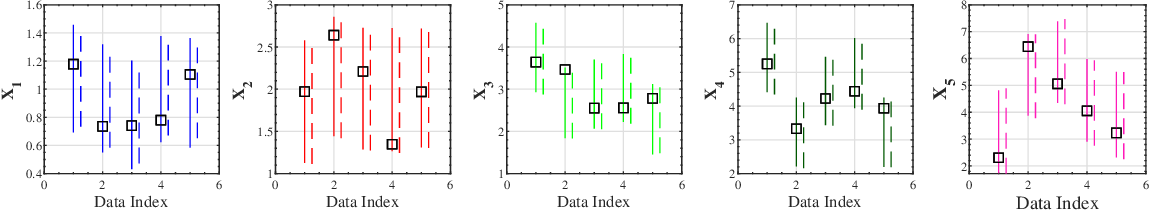}
	\caption{Results obtained using the code generated by ChatGPT for Bayesian updating on a 5 dimensional data with missing values. Square denotes the true value, solid line and dotted line denote the $95\%$ confidence interval obtained using ChatGPT, and the benchmark method \citep{ching2019constructing} respectively.}
	\label{Bayes_3}
\end{figure}

 Now, let us check the correlation coefficients. The results are presented in Fig. \ref{Bayes_1}. The upper triangular matrix presents the histograms of the obtained estimates, and the solid vertical line denotes the actual value used to sample the data. It is evident that correlation can also be modelled satisfactorily. Finally, using all the sampled parameters, posterior predictive samples are obtained, and the empirical contours for all the pairs (total 10) along with the measured data are presented in the lower triangular matrix of Fig. \ref{Bayes_1}. Although the data was complete in this case, let us check the capability of the code to predict the missing values. In each dimension, five known values are removed, and the estimates using the obtained code are presented in Fig. \ref{Bayes_3}. The results are also independently benchmarked against the method presented in \cite{ching2019constructing}. (the code was provided in personal communication). Remarkably, the estimates are almost similar. Overall, the results obtained in Fig. \ref{Bayes_2}, \ref{Bayes_1}, and \ref{Bayes_3} suggested that the code generated by ChatGPT seems to work correctly.

\begin{samepage}

	\begin{tcolorbox}[human, title=Human:]
write me a matlab code for estimating joint distribution paramters using Bayesian Gibbs sampling from a m rows n columns data , where each data point is iid and columns are cross correlated, also it should be able to handle missing value and iteratively predict the missing values based on their condtional densities.
	\end{tcolorbox}
	
	\begin{tcolorbox}[assistant, title=AI Assistant:]
Sure, here is a MATLAB code that uses Bayesian Gibbs Sampling to estimate joint distribution parameters for a dataset with missing values. The code assumes that the data follows a multivariate normal distribution and uses the Gibbs Sampling method to handle missing values and iteratively predict them based on their conditional densities.
	\end{tcolorbox}
	
	\begin{tcolorbox}[code, title=Matlab Script]
		\begin{lstlisting}[caption={Bayesian updating},label={4th:label},lineskip=-2pt,language=Matlab]
function [mu_samples, Sigma_samples, imputed_data_samples] = bayesian_gibbs_sampling(data, num_iterations, burn_in)
% Bayesian Gibbs Sampling for estimating joint distribution parameters
% and imputing missing values.
%
% Input:
%   - data: m x n matrix with missing values represented as NaN
%   - num_iterations: Number of iterations for Gibbs Sampling
%   - burn_in: Number of iterations to discard as burn-in
%
% Output:
%   - mu_samples: Samples of the mean vector (num_iterations - burn_in) x n
%   - Sigma_samples: Samples of the covariance matrix (num_iterations - burn_in) x n x n
%   - imputed_data_samples: Imputed data matrix for each iteration m x n x (num_iterations - burn_in)

% Dimensions
[m, n] = size(data);

% Priors
mu_prior_mean = zeros(1, n);      % Non-informative prior mean
mu_prior_cov = 100 * eye(n);      % Non-informative prior covariance (large value)
nu_prior = n + 2;                 % Degrees of freedom for the Inverse-Wishart distribution
Psi_prior = eye(n);               % Scale matrix for the Inverse-Wishart distribution

% Initialize parameters
mu_est = nanmean(data);           % Initial mean estimate (ignoring NaNs)
Sigma_est = nancov(data);         % Initial covariance estimate (ignoring NaNs)

% Initialize missing values randomly
imputed_data = data;
missing_idx = isnan(data);
imputed_data(missing_idx) = randn(sum(missing_idx(:)), 1);

% Storage for samples
mu_samples = zeros(num_iterations - burn_in, n);
Sigma_samples = zeros(num_iterations - burn_in, n, n);
imputed_data_samples = zeros(m, n, num_iterations - burn_in);

% Gibbs Sampling
for iter = 1:num_iterations
% Sample mu given Sigma and data
Sigma_n = inv(inv(mu_prior_cov) + m * inv(Sigma_est));
mu_n = Sigma_n * (inv(mu_prior_cov) * mu_prior_mean' + m * inv(Sigma_est) * mean(imputed_data)');
mu_est = mvnrnd(mu_n, Sigma_n);

% Sample Sigma given mu and data
Psi_n = Psi_prior + (imputed_data - mu_est)' * (imputed_data - mu_est);
Sigma_est = iwishrnd(Psi_n, nu_prior + m);

% Update missing values
for i = 1:m
observed_idx = ~isnan(data(i, :));
missing_idx_row = isnan(data(i, :));

if any(missing_idx_row)
mu_obs = mu_est(observed_idx);
mu_mis = mu_est(missing_idx_row);

Sigma_obs = Sigma_est(observed_idx, observed_idx);
Sigma_mis = Sigma_est(missing_idx_row, missing_idx_row);
Sigma_obs_mis = Sigma_est(observed_idx, missing_idx_row);

observed_data = imputed_data(i, observed_idx);

% Conditional mean and covariance
cond_mean = mu_mis + Sigma_obs_mis' / Sigma_obs * (observed_data' - mu_obs');
cond_cov = Sigma_mis - Sigma_obs_mis' / Sigma_obs * Sigma_obs_mis;

% Sample from conditional distribution
imputed_data(i, missing_idx_row) = mvnrnd(cond_mean, cond_cov, 1);
end
end

% Store samples after burn-in period
if iter > burn_in
mu_samples(iter - burn_in, :) = mu_est;
Sigma_samples(iter - burn_in, :, :) = Sigma_est;
imputed_data_samples(:, :, iter - burn_in) = imputed_data;
end
end
end


			
		\end{lstlisting}
	\end{tcolorbox}
\end{samepage}

\section*{Discussion: limitations, challenges, and future scope}

The previous results only showed the successful results. Even with the successful results, the authors encountered difficulty generating FORM codes. The query and response of a failed example is demonstrated in Listing \ref{5th:label} (Appendix section). The $\beta=0$ and $P_f=0.5$ is obtained on running this code. $P_f$ calculated using MCS with $1e8$ samples for the same problem is $0.18$. To check further we changed the $\mu=[5,3]$ in the original response to $\mu=[7,3]$. The revised results for $P_f$ and $\beta$ are same using the code, i.e., the results do not change, suggesting incorrect code. The MCS $P_f$ for the changed mean is $0.0035$, confirming that the code indeed is incorrect. Although we tried to improve this code by providing the potential issue to the Chat-GPT, the code barely improved and still gave us incorrect results. We got the correct code only when we included the "constrained optimization" keyword in a fresh prompt. One suggestion for future users is if, upon numerous iterative (let's say 5) prompts, the codes are not improving, it is better to start a fresh conversation with a new prompt to generate the results. In some cases, the LLM may perform poorly simply because of vague or imprecise instructions. Therefore, a concise description of the problem in the prompt is helpful.

Subset simulation codes and Bayesian updating codes were generated correctly in a single attempt. For random field simulation, when we asked it to choose its method for simulating random fields, it decided on an FFT-based method, saying that it is efficient (which is not incorrect). The query and response are presented in Listing \ref{6th:label} (Appendix section). It could generate random fields, but the authors could not satisfactorily conclude if the the results were correct (see Fig. \ref{approach0 } in Appendix). Therefore, we revised the prompt to use the covariance matrix decomposition method with separate correlation assumptions (see Listing \ref{3rd:label}). Note that this study didn't use any tuning techniques or advanced prompting strategies but just plain vanilla zero-shot prompting because that's what we expect from an average practitioner's point of view.

Another challenge we found using the codes was the time and effort spent verifying whether the code generated the correct results. Although one of the purposes of generative AI is to aid people who are not experts in the domain of risk and reliability, the codes still need to be verified. This sounds like a paradox as it again requires expert risk and reliability knowledge. Although code verification is difficult, verifying the results on some simple problems or using other benchmark methods is relatively less difficult. This aspect also means that the task of generating codes based on the current state of the art of generative AI cannot be fully automated; it can only be semi-automated as some domain knowledge is still needed to verify the code and results. Nevertheless, it lowers the expert knowledge barrier and helps accelerate the programming of risk and reliability tasks.

Beyond mere efficiency gains in programming reliability algorithms, LLMs can potentially contribute to the improvement of code quality. In future, LLMs trained on vast repositories of high-quality reliability codes can implicitly learn best practices. These learned best practices can result in more robust solutions when applied to reliability engineering. For instance, LLMs can assist in generating comprehensive documentation for complex algorithms, making it easier for other researchers and practitioners to understand and build upon existing work. This improved knowledge transfer is particularly crucial in the field of civil engineering, where project-specific adaptations of risk assessment methods are often necessary.

	\section*{Conclusions}
	
This study explored the possibility of using generative AI to program reliability algorithms. Specifically, ChatGPT was utilized to generate MATLAB codes for four classical reliability methods, i.e., FORM, Subset simulation, Random field simulation, and Bayesian updating. The results obtained for the FORM and Subset simulation were verified using Monte Carlo simulation. For the random field simulation, the generated random field was verified to check the correct specification of correlation length and x and y direction and an already public random field generator code was used in the literature. The Bayesian updating results were verified using numerical examples and an independently available code. The code for Subset simulation and Bayesian updating using the Gibbs sampler were generated in a single attempt correctly. The random field simulation codes and FORM codes required fresh and different prompts after some failures ( using other prompts) to generate the correct code. 

Overall, the results are promising, but still, in some cases, they require a lot of efforts to generate correct codes in addition to the verification of the code. This aspect implies that some domain knowledge is still needed to verify the code and results. Nevertheless, it lowers the knowledge barrier and also helps accelerate the programming of reliability algorithms. This study used only ChatGPT to test the capability for code generation of reliability methods since it is the most popular. There are many recent generative AI models, such as Claude \citep{anthropic2023claude},  Gemini \citep{google2023gemini}, Llama \citep{meta2023llama}, SenseChat \citep{SenseTime_2024}, Ernie Bot \citep{Baidu_2024}, and Grok \citep{tencent2023grok},  which can be explored for further programming tasks in reliability.

	\section*{Declaration of competing interest}
	The authors declare that they have no known competing financial interests or personal relationships that could have appeared to influence the work reported in this paper.
	
	\section*{\textcolor{black}{Acknowledgment}}
	
	This research was substantially supported by the Shanghai Municipal Human Resource and Social Security Bureau (2023554), the National Key Research and Development Program of China (2021YFB2600500), the National Natural Science Foundation of China (42072302, 52025094), the Key Innovation Team Program of MOST of China (2016RA4059) and Fundamental Research Funds for the Central Universities. Finally, the use of Chat-GPT 4.0 for generating the codes in this study is gratefully acknowledged. 
	
	\section*{Data availability}
	All the data or computer codes that support the findings of this study are available in this study.

	\section*{Appendix}
	\appendix
	This appendix contains Listing \ref{5th:label}, Listing \ref{6th:label}, Fig. \ref{approach0 }, and Fig. \ref{approach1}. Listing \ref{5th:label} is for a failed FORM example (incorrect results) and Listing \ref{6th:label} is for an unused random field simulation example.
\renewcommand{\thefigure}{A\arabic{figure}}

\setcounter{figure}{0}

	\begin{samepage}

		\begin{tcolorbox}[human, title=Human:]
			write me a matlab code for advanced first oder reliability method using some non linear limit state function
		\end{tcolorbox}
		
		\begin{tcolorbox}[assistant, title=AI Assistant:]
			
			The Advanced First Order Reliability Method (AFORM) extends the basic FORM approach by including nonlinear transformations between the original space (X-space) and the standard normal space (U-space). In AFORM, the iterative process finds the design point (most probable point) more accurately by accounting for nonlinearity in the limit state function and possibly using improved search directions in the transformed space.
			
			Here’s a MATLAB code implementing AFORM using a nonlinear limit state function:
			
		\end{tcolorbox}
		
		\begin{tcolorbox}[code, title=Matlab Script]
			\begin{lstlisting}[caption={Failed Example (FORM)},label={5th:label},lineskip=-2pt,language=Matlab]
% Advanced First Order Reliability Method (AFORM) for a Nonlinear Limit State Function

clear;clc;

% Define the mean and standard deviation of the random variables
mu = [5; 3];              % Mean values for x1 and x2
sigma = [1; 1];           % Standard deviations for x1 and x2

% Define the initial guess for the design point in the original space
x = mu;

% Convergence tolerance and maximum number of iterations
tolerance = 1e-6;
max_iterations = 100;

% Initial guess for the reliability index beta
beta = 1;

% Main iteration loop for AFORM
for iter = 1:max_iterations
% Calculate the limit state function value at the current design point
g_x = x(1)^2 + x(2)^2 - 25;

% Calculate the gradient of the limit state function with respect to x1 and x2
grad_g = [2 * x(1); 2 * x(2)];

% Calculate the alpha vector (normalized gradient in U-space)
alpha = grad_g ./ (sigma.^2);  % Adjust for non-normal variables
alpha = alpha / norm(alpha);

% Calculate the new beta value using the improved transformation
u = (x - mu) ./ sigma;
beta_new = alpha' * u;

% Update the design point in the U-space
u_new = -beta_new * alpha;

% Transform back to the original space
x_new = mu + sigma .* u_new;

% Check for convergence
if abs(beta_new - beta) < tolerance
beta = beta_new;
x = x_new;
break;
end

% Update beta and design point for the next iteration
beta = beta_new;
x = x_new;
end

% Calculate the probability of failure
Pf = normcdf(-beta);

% Display the results
fprintf('Converged Reliability Index (beta): %.4f\n', beta);
fprintf('Probability of Failure: %.6e\n', Pf);
fprintf('Design Point (in original space):\n');
fprintf('  x1* = %.4f\n', x(1));
fprintf('  x2* = %.4f\n', x(2));
			\end{lstlisting}
		\end{tcolorbox}
	\end{samepage}

		\begin{samepage}

		\begin{tcolorbox}[human, title=Human:]
write me a matlab code for generating 2D random field, use a method as per your choice
		\end{tcolorbox}
		
		\begin{tcolorbox}[assistant, title=AI Assistant:]
			
Here's a MATLAB code that generates a 2D random field using the Fast Fourier Transform (FFT) method. This method is efficient and widely used for generating random fields with a specified correlation structure.
			
		\end{tcolorbox}
		
		\begin{tcolorbox}[code, title=Matlab Script]
			\begin{lstlisting}[caption={Unused Example (Random Field Simulation)},label={6th:label},lineskip=-2pt,language=Matlab]
% 2D Random Field Generation using FFT

clear;clc;

% Parameters for the grid
nx = 256; % Number of grid points in x direction
ny = 256; % Number of grid points in y direction
Lx = 100; % Domain size in x direction
Ly = 100; % Domain size in y direction
lx = 10;  % Correlation length in x direction
ly = 5;   % Correlation length in y direction

% Generate grid spacing
dx = Lx / nx;
dy = Ly / ny;
[x, y] = meshgrid(0:dx:Lx-dx, 0:dy:Ly-dy);

% Generate wavenumbers
kx = 2 * pi * [0:nx/2 -nx/2+1:-1] / Lx;
ky = 2 * pi * [0:ny/2 -ny/2+1:-1] / Ly;
[KX, KY] = meshgrid(kx, ky);

% Define the power spectrum (Fourier transform of the correlation function)
% Assuming an exponential correlation function: exp(-sqrt((x/lx)^2 + (y/ly)^2))
S = exp(-sqrt((KX * lx).^2 + (KY * ly).^2));

% Generate a complex Gaussian random field in the Fourier space
Z = sqrt(S) .* (randn(ny, nx) + 1i * randn(ny, nx));

% Apply inverse FFT to get the random field in real space
field = real(ifft2(Z));

% Normalize the field to have zero mean and unit variance
field = (field - mean(field(:))) / std(field(:));

% Plot the 2D random field
figure;
imagesc(x(1, :), y(:, 1), field);
axis equal;
colorbar;
title('2D Random Field Generated Using FFT');
xlabel('x');
ylabel('y');

				
			\end{lstlisting}
		\end{tcolorbox}
	\end{samepage}
	
	\begin{figure}[htp]
	\centering
	\includegraphics[trim=0cm 0cm 0cm 0cm, clip=true,width=0.75\textwidth]{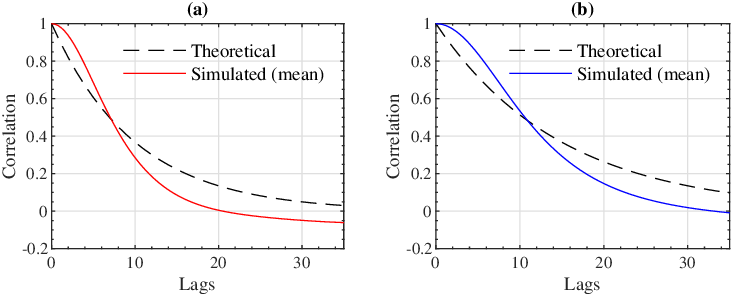}
	\caption{Theoretical and simulated correlation for the FFT based random field code generated by ChatGPT (Listing \ref{6th:label}, unused) in: (a) x-direction; and (b) y-direction. Mean is computed over 1000 realizations. }
	\label{approach0 }
\end{figure}

	\begin{figure}[htp]
	\centering
	\includegraphics[trim=0cm 0cm 0cm 0cm, clip=true,width=0.75\textwidth]{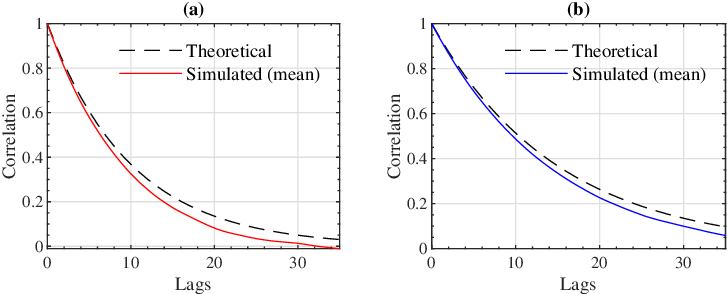}
	\caption{Theoretical and simulated correlation for the covariance matrix decomposition based random field code generated by ChatGPT (Listing \ref{3rd:label}) in: (a) x-direction; and (b) y-direction. Mean is computed over 1000 realizations. }
	\label{approach1}
\end{figure}

	\bibliographystyle{apalike}
	\bibliography{Manuscript_R0}

\end{document}